\documentclass[aps,amsmath,amssymb,showpacs]{revtex4}
\usepackage{color}
\usepackage{graphicx}

\begin{document}

\title{Von Neumann's No Hidden Variable Theorem }

\author{M. Revzen}
\affiliation {Department of Physics, Technion - Israel Institute of Technology,
Haifa 32000, Israel}

\date{\today}

\begin{abstract}

Von Neumann use 4 assumptions to derive the Hilbert space (HS) formulation of quantum mechanics (QM). Within this theory dispersion free ensembles do not exist. To accommodate a theory of quantum mechanics that allow dispersion free ensemble some of the assumptions need be modified. An existing formulation of QM, the phase space (PS) formulation allow dispersion free ensembles and thus is qualifies as an hidden variable theory.  Within the PS  theory we identify the violated assumption (dubbed I in the text) to be the one that requires that the value r for the quantity $\mathbb{R}$  implies the value f(r) for the quantity $f(\mathbb{R})$. We note that this violation arise due to tracking within c-number hidden variable theory of the operator ordering involved in HS theory as is required for a 1-1 correspondence between the theories.

\end{abstract}

\pacs{XXX}

\maketitle

\section{Introduction}

In Chapter III of his definitive monograph, Mathematical Foundation of Quantum Mechanics, von Neumann (VN) reduced all assertions of quantum mechanics
(QM) to a statistical formula \cite{vn} p.295. He then, in Chapter IV, derived this formula from a "few  qualitative assumptions" which allowed
him to check "the entire  structure of QM". These are distilled into \cite{ms}  four assumptions, two, dubbed A' and B', deal with physical quantities (observables) and their measurements and two dubbed I and II relates to mathematical structure (i.e. to the Hilbert space (HS) framework) of the theory. Assumption A' is \cite{ms} : There exists an expectation function Exp from physical quantities $\mathbb{R}$ to real numbers r. "With each quantity we include the directions as to how it is to be measured" \cite{vn} p.297. E.g. if $\mathbb{R}$ is a (physical) quantity that one knows how to measure and f(x) is some real function then the quantity $f(\mathbb{R})$ is measured by measuring $\mathbb{R}$ and apply f to the outcome of the $\mathbb{R}$ measurement. Assumption B' stipulates that a complex physical quantity expressible as a linear combination of several physical quantities $\mathbb{R}, \mathbb{S},..$ not necessarily simultaneously measurable abides by

$$Exp(a\mathbb{R}+b\mathbb{S}+...)\;=\;aExp(\mathbb{R})+bExp(\mathbb{S})+ ...$$

a, b,... are real numbers. Assumption I \cite{bub}: If the quantity $\mathbb{R}$  has the operator $\hat{R}$ then the quantity $f(\mathbb{R})$ has the operator $f(\hat{R})$. Assumption II \cite{bub} If the quantities $\mathbb{R},\mathbb{S}...$ has operators $\hat{R}, \hat{S}, ...$ then the
the quantity $(\mathbb{R}+\mathbb{S}+...)$ has the operator $(\hat{R}+\hat{S} +..)$.\\

It is important to stress the distinction between A' and I. A' explains the notion simultaneous measurements while I relates to functional relations between two quantities.\\

The 4 assumptions allow VN to prove \cite{vn} p. 316, that the expectation function of the quantity $\mathbb{R}$ is uniquely defined by the trace function,

\begin{equation}\label{tr}
 Exp(\mathbb{R}) \;=\;tr\bar{U}\hat{R},
 \end{equation}

where $\overline{U}$ represents and Hermitian non negative operator. It is the statistical operator representing the physical system
under study and is independent of the quantity measured. Thus the attributes of the statistical operator provides "insight on the entire structure of QM" (p.295), i.e. the allowed ensembles. VN considers two qualities of general importance: (1) A dispersion free ensemble, and (2) an homogeneous ensemble.\\

Dispersion in observation of physical quantity is

\begin{equation}\label{dis}
Dis(\mathbb{A}) \;=\; [Exp(\mathbb{A})^2]\;-\;[Exp(\mathbb{A})]^2\;= \;tr(\overline{U}\hat{A}^2)\;-\;[tr(\overline{U}\hat{A}]^2.
\end{equation}

Thus a dispersion free ensemble is $\overline{U}$ for which $tr(\overline{U}\hat{A}^2)\;=\;[tr(\overline{U}\hat{A}]^2 \;\;\forall\;\hat{A}.$
VN proved (p.321) as is briefly discussed below, that dispersion free ensembles do not exist within his 4 assumptions.\\

An homogeneous ensemble \cite{bub}  is such that for all $\mathbb{A}$ "$Exp(\mathbb{A})$ can not be expressed as a convex combination of two different expectation value functions neither of which equal to $Exp(\mathbb{A})$ ". I.e. the normalized $\overline{U}$ is the same for all three.
VN proved \cite{vn} p. 323, that such ensembles are necessarily projectors: $\overline{U}\;=\;|\phi><\phi|$ for some $|\phi>$. We note that these ensembles are dispersive.\\

VN investigates the proposition of adjoining new parameters $\lambda = {\lambda_1,\lambda_2,...}$, which are  "hidden variables" (hv) that delineate
among members of an ensemble of homogeneous states $|\phi> \rightarrow |\phi,\lambda>$  with the projector $|\phi,\lambda><\lambda,\phi|$ dispersion free, and the homogeneous
ensemble {$|\phi><\phi|$} some sort of an average over $\lambda$. A hv theory is a theory that allow dispersion free ensembles. The adjointly parametrized system would be a hidden variable (hv) theory. VN proved that within his 4 assumptions it is impossible to embed QM within a a finer grained hv theory. Thus the very same assumptions that allow the derivation of central quantum mechanical statistical formula precludes direct embedment of the theory within an hv theory: to formulate QM as a hv theory some of the assumption would have to be abandoned/modified.\\

Conversely an existing  hv theory of QM would have to violate/modify some of the 4 assumptions. In such case,  VN noted, the violation would undermine the fundamental structure of the hitherto successful HS formulation. In the next section we argue that such hv theory does exist - it is the
well known PS formulation QM \cite{zachos,curtright} we then identify the violated assumptions and note that indeed it (the PS formulation) is fundamentally different from the HS formulation.\\

\section{Hidden variable theory}

VN identifies assumption I and or II as those that are to be modified/abandoned if we wish to allow the theory to be 'embeded' within a hv theory \cite{vn} (p.324) (see also \cite{bub},\cite{dieks},\cite{ms}). Modifications of these assumptions entails a major change in the formalism.\\

The abandonment/modification of assumption(s) I/II  to allow possible formulation of an hv theory for QM acquired
a controversial status with the publication of Bell \cite{bell1} (and earlier \cite{hermann}). Bell contends that VN theory is of limited value
in that it assigns the value for the ill defined sum of physical quantities $\mathbb{A} + \mathbb{B}$ as the sum of the assigned values of $\mathbb{A}$ and $\mathbb{B}$. What in effect is Bell's  contention is that VN set of 4 assumptions that purports to underpin QM do not form a tight set - assumption B' is redundant \cite{ms} and  it should be abandoned to allow  him to consider a more meaningful no hv formulation (of QM) theory. His criticism enjoyed considerable support  e.g. \cite{mermin,haag}. Lately J. Bub \cite{bub} and D. Dieks \cite{dieks} challenged this criticism and most recently \cite{ms} reaffirms Bell's criticism of von Neumann no hidden variable theory.\\

VN terminology is somewhat out of date and hence the interpretation of his argument may be controversial. However since we have
an intuitively appealing hv theory in the PS formulation of QM it is possible to explicitly spot which of VN assumptions were
abandoned within the PS space formulation. We contend that it assumption I, in the form \cite{ms} "For any real-valued function f, if the quantity $\mathbb{A}$ has an operator $\hat{A}$
then the quantity $f(\mathbb{A})$ has the operator $f(\hat{A})$" that was abandoned assumption accompanied indeed with major change in the formalism. Thus the assumption

\begin{equation}\label{xfx}
\mathbb{A}\; \leftrightarrow \;\hat{A}\;\;\Rightarrow\;\;f(\mathbb{A})\;\leftrightarrow\;f(\hat{A}),
\end{equation}

does not apply when Phase Space variables replaces those of Hilbert Space in accounting for QM.
To demonstrate this we recall that the PS expression for physical quantity is the dequantized Weyl transform \cite{w,wq}, of its Hilbert space expression
Eq.(\ref{tr}). Consider the operator $\hat{H}=\frac{1}{2}(\hat{q}^2 + \hat{p}^2)$ and $f(\hat{H})=\hat{H}^2$. Denoting the Weyl transform by tilde
we have

\begin{eqnarray}
\tilde{H}&=&\frac{1}{2}(q^2 +p^2)= H, \nonumber \\
\tilde{H^2}&=&\frac{1}{4}(q^4+p^4 +2q^2p^2 -1)=H^2-1/4
\end{eqnarray}

The correction term (1/4) is due to the Weyl ordering role in the Weyl transform. This illustrates that within the PS formulation
the correspondence $\mathbb{A}$ to A does not imply the correspondence $f(\mathbb{A})$ to $f(A)$ because $(\tilde{f})(A)$ is not equal
to $f(\tilde{A})$ due to involvement of non co measureable elements. Thus here the hv variable theory involves the abandonment of
assumption I in compliance with VN analysis. It is indeed associated with a mathematical formulation (of QM) quite distinct from the HS one - the PS formulation.\\

Thus whereas  within the HS formulation the association of the physical quantity $\mathbb{R}$  with the operator $\hat{R}$  implies the association of the quantity  $f(\mathbb{R})$ with $f(\hat{R})$ - within the PS formulation of QM the association $\mathbb{R}$ with $\tilde{R}$
(the Weyl transform of $\hat{R}$) implies the association of $f(\mathbb{R})$  with $\tilde{f}(\hat{R})$ which is unequal to $f(\tilde{R})$ in general: Thus for
$\hat{R}=\hat{A}+\hat{B}$ the linear term is $\tilde{\hat{R}}$ however, for $[\hat{A}, \hat{B}]_- \neq 0,$ the non linear $\tilde{f}(\hat{R})\neq f(\tilde{R})$ as is illustrated above for $f(x)=x^2$.\\ 

We conclude thence that assumption I is abandoned  to allow the Weyl transform  map the HS formulation of QM to a hv theory validating thereby the contention of \cite{bub, dieks}. The formalism is entirely different e.g. the adjoining of two quantities is now via the so called star $\star$ product rather than standard multiplication much as expected by VN. \\

\section {Concluding Remarks}

The assumptions introduced by von Neumann to allow the derivation of the central quantum mechanical statistical formula implied that
dispersion free ensemble does not exist. To embed the theory within a hidden variable theory i.e. a theory that allow dispersion free ensembles
requires modification of some assumptions. The interpretation to the modification put forward by von Neumann
were criticized and led to some controversy. These are resolved by studying the phase space formulation of quantum mechanics which is identified as a hidden variable theory.\\

The Phase space formulation of QM is recognized as a hidden variable theory since it allows dispersion free ensembles. Direct comparison
of the theory with the canonical Hilbert space formulation show that the requirement for having for any physical quantity $\mathbb{R}$ whose operator representative is $\hat{R}$  imply that the representative of the quantity $f(\mathbb{R})$, for every real function f,  is
$f(\hat{R})$ does not carry over to the embeding \cite{klein} hidden variable theory. The violation comes about because  the Weyl transformation \cite{wq} that assures the 1-1 correspondence with the Hilbert space formulation and PS is such that $\tilde{f}(R)\neq f(\tilde{R})$, i.e. the Weyl transform of the function of composite variable R is unequal, in general, to the function of the Weyl transform of R.\\

\end{document}